\documentclass{aa}

\usepackage[final]{graphicx}
\graphicspath{{./figures/}}

\usepackage[fleqn]{amsmath}
\usepackage{amstext,amssymb}
\usepackage{natbib}
\bibpunct{(}{)}{;}{a}{}{,}

\newcommand{\strong}[1]{\ifmmode\mathbf{#1}\else\textbf{#1}\fi}
\usepackage{color}
\newcommand{\microns}{\ensuremath{\mu\text{m}}}

\newcommand{\Rsun}{\ensuremath{R_\odot}}
\newcommand{\Lsun}{\ensuremath{L_\odot}}
\newcommand{\Msun}{\ensuremath{M_\odot}}
\newcommand{\Mjup}{\ensuremath{M_{\rm Jupiter}}}
\newcommand{\Mdot}{\ensuremath{\dot M}}
\newcommand{\Mstar}{\ensuremath{M_\star}}
\newcommand{\AU}{\ensuremath{\text{AU}}}
\newcommand{\degree}{\ensuremath{\text{deg}}}
\newcommand{\yr}{\ensuremath{\text{yr}}}
\renewcommand{\arcsec}{\ensuremath{\text{arcsec}}}

\def\Tdisk{\ensuremath{T_0}}
\def\rdisk{\ensuremath{r_0}}
\def\rmin{\ensuremath{r_\text{min}}}
\def\rmax{\ensuremath{r_\text{max}}}
\def\qdisk{\ensuremath{q}}
\def\idisk{\ensuremath{i}}
\def\thetadisk{\ensuremath{\theta}}

\def\dsep{\ensuremath{d_\mathrm{spot}}}
\def\thetasep{\ensuremath{\theta_\mathrm{spot}}}
\def\Tspot{\ensuremath{T_\mathrm{spot}}}
\def\rspot{\ensuremath{r_\mathrm{spot}}}

\def\Bp{\ensuremath{B_\text{p}}}
\def\vecBp{\ensuremath{\vec{B}_\text{p}}}

\def\flux{\ensuremath{F_\lambda}}
\def\vis{\ensuremath{V_\lambda}}

\newcommand\idiff[2][]{\ensuremath{\,\text{d}^{#1}#2}}
\newcommand\integ[4]{\int_{#1}^{#2} #3 \idiff{#4}}

\renewcommand{\vec}[1]{\ensuremath{\mathbf{#1}}}

\begin{document}
\title{New insights on the AU-scale circumstellar structure of FU Orionis}
\author{
       F. Malbet \inst{1}
  \and R. Lachaume \inst{2}
  \and J.-P. Berger\inst{1}
  \and M. Colavita\inst{3}
  \and E. Di Folco\inst{4}
  \and J.A. Eisner\inst{5}
  \and B.F. Lane\inst{6}
  \and R. Millan-Gabet\inst{7}
  \and D. S\'egransan \inst{8}
  \and W. Traub \inst{9}
}

\institute{ 
  Laboratoire d'Astrophysique de Grenoble, UMR CNRS/UJF 5571, BP 53,
  F-38041 Grenoble cedex 9, France
  \and Max-Planck-Institut f\"ur Radioastronomie, Auf dem H\"ugel 69,
  D-53121 Bonn, Germany
  \and Jet Propulsion Laboratory, California Institute of Technology,
  4800 Oak Grove Dr., Pasadena, CA 91109, USA
  \and European Southern Observatory, Karl Schwarzschild Strasse 2,
  D-85748 Garching, Germany
  \and Department of Astronomy, California Institute of Technology, MC
  105-24, Pasadena, CA 91125
  \and Center for Space Research, Massachusetts Institute of
  Technology, 70 Vassar Street, Cambridge, MA 02139, USA
  \and Michelson Science Center, California Institute of Technology,
  770 S. Wilson Ave., Pasadena, CA 91125, USA
  \and Observatoire de Gen\`eve, 1290 Sauverny, Switzerland
  \and Harvard-Smithsonian Center for Astrophysics, 60 Garden St.,
  Cambridge, MA 02138, USA
}

\offprints{F. Malbet}
\mail{Fabien.Malbet@obs.ujf-grenoble.fr}
\date{Received dat: 17 December 2004 / Accepted date: 23 March 2005}

\authorrunning{F. Malbet et al.}
\titlerunning{New insights on the AU-scale circumstellar structure of FU Orionis}

\abstract{ We report new near-infrared, long-baseline interferometric
  observations at the AU scale of the pre-main-sequence star FU
  Orionis with the PTI, IOTA and VLTI interferometers. This young
  stellar object has been observed on 42 nights over a period of 6
  years from 1998 to 2003. We have obtained 287 independent
  measurements of the fringe visibility with 6 different baselines
  ranging from 20 to 110 meters in length, in
  the $H$ and $K$ bands. Our data resolves FU Ori 
  at the AU scale, and provides new constraints at shorter
  baselines and shorter wavelengths. Our extensive $(u,v)$-plane
  coverage, coupled with the published spectral energy distribution
  data, allows us to test the accretion disk scenario. We find that the
  most probable explanation for these observations is that FU Ori
  hosts an active accretion disk whose temperature law is consistent
  with standard models and with an accretion rate of $\Mdot = (6.3 \pm
  0.6) \times 10^{-5} (\Mstar/\Msun)^{-1} \, \Msun/\yr$.  We are able
  to constrain the geometry of the disk, including an inclination of
  $55_{-7}^{+5}\,\degree$ and a position angle of
  $47_{-11}^{\hphantom{0}+7}\,\degree$.  In addition, a 10 percent
  peak-to-peak oscillation is detected in the data (at the two-sigma
  level) from the longest baselines, which we interpret as a possible
  disk hot-spot or companion. The still somewhat limited $(u, v)$
  sampling and substantial measurement uncertainty prevent us from
  constraining the location of the spot with confidence, since many
  solutions yield a statistically acceptable fit.  However, the
  oscillation in our best data set is best explained with an
  unresolved spot located at a projected distance of $10\pm1\,\AU$ at
  the $130\pm1\,\degree$ position angle and with a magnitude
  difference of $\Delta K \approx 3.9\pm0.2$ and $\Delta H \approx
  3.6\pm0.2$\,mag moving away from the center at a rate of
  $1.2\pm0.6\,\AU\,\yr^{-1}$.  Although this bright spot on the
  surface of the disk could be tracing some thermal instabilities in
  the disk, we propose to interpret this spot as the signature of a
  companion of the central FU Ori system on an extremely eccentric
  orbit. We speculate that the close encounter of this
  putative companion and the central star could be the explanation of
  the initial photometric rise of the luminosity of this object.
}
\maketitle

\section{Introduction}

FU Orionis is a variable stellar system representative of a small
class of pre-main-sequence (PMS) objects named FUors \citep{Amb71,
Amb82}.  These stars show a rise of their luminosity over a
timescale of a few hundred years. Some FUors, observed before their flare-ups,
display the spectral characteristics of a typical T Tauri star
(TTS). It is now widely accepted that most TTS go through this type of
very short FUor-like outburst phase, possibly several times during
the early stages of their existence \citep{Her1977,Ken2000}.

FUors display other interesting observational peculiarities: an
infrared excess that is far larger than the one found in T
Tauri stars, and very broad absorption lines that appear to be double.
The exact nature of FUors properties is still controversial, 
with essentially two competing explanations:
\begin{enumerate}
\item {\bf Rapidly Rotating G Supergiant:} \citet{PH92,Her03} have
  shown that an FUori optical spectrum can be reproduced by a G-type
  supergiant chromosphere overlaid with a rising, cooler absorbing
  shell. As proposed by \citet{Lar80}, the star is rotating near 
  breakup, and could be responsible for the flare-up.

\item {\bf Accretion-disk Instability:} \citet{HK85,HK96} have proposed
  that the flare-up is a phenomenon not of the star itself but rather
  the result of a major increase in the surface brightness of the
  circumstellar accretion disk surrounding a young T Tau-type star.
  Since the idea was first introduced, theories of instabilities
  intrinsic to such an accretion disk have been examined by
  \citet{CLP90,Bell94,Bell99} and \citet{KL99}. A related scenario is
  the passage of a companion star close to the primary star, which may
  cause an increase in the accretion rate \citep{BB92, CS96}.
\end{enumerate}

\citet{HK96} have three main arguments to explain the origin of
spectroscopic properties of FUors with the disk accretion
model. First, the separation of the peaks in the double absorption
lines decreases at longer wavelengths. Such behavior is expected in a
Keplerian accretion disk, since the outer parts rotate slower than the
inner ones. Second, the presence of CO bands requires environmental
conditions cooler than the stellar surface. Such conditions can be
found in the inner regions of the disk. Third, as is the case for
other young stellar objects (YSOs), the infrared excess produced by a
disk can explain the shape of the observed FUor infrared spectral
energy distributions (SEDs). However \citet{Her03} claim that ``the
spectroscopic properties of FUors that have been argued as proof of
the accretion disk picture [...]  are found in a number of much older
high-luminosity stars where there is no reason to think that an
accretion disk is present''. \citet{HHC04}, based on additional
optical spectroscopic observations of FUors, point out that the
chromospheric emission reversal of lines with very different strengths
and excitation potentials required in the framework of a rapidly
rotating star, as well as the decrease of the rotational broadening
with wavelength, do not favor the chromosphere model.

High angular resolution observations of these objects can bring new
observational constraints and help to solve this controversy. FU Ori
is located in Orion ($d=450\mbox{ pc}$; 1 AU corresponds to 2.2 mas.) Recent
observations with adaptive systems \citet{WAHP04} showed that FU Ori
has a companion located $0.5\,\arcsec$ south with a magnitude difference of
about 4 magnitudes in the near infrared. This companion has
been found to also be a young star \cite{RA2004}.

The first high angular resultion observations of PMS stars with an
infrared interferometer were carried out on FU Ori itself \citep[][,
hereafter paper I]{Mal98}, one of the brightest YSOs, using the
\emph{Palomar Testbed Interferometer} (PTI). Unfortunately the limited
number of baselines was not sufficient to definitively conclude on the
exact nature of FU Orionis. \citet{Ber00} published additional
visibility data on FU Orionis obtained with PTI and the \emph{Infrared
  and Optical Telescope Array} (IOTA).

In this paper, we present the results of a 6 year campaign on FU Ori
with the most advanced infrared interferometers around the world,
including the recently available \emph{Very Large Telescope
  Interferometer} (VLTI).  Section~\ref{sect:obs} summarizes the observations;
Sect.~\ref{sect:dataproc} describes briefly the data processing and
Sect.~\ref{sect:results} presents the results. We analyze and
interpret these new observations in Sect.~\ref{sect:interp} and we
discuss them in the framework of disk models and star
formation theories in Sect.~\ref{sect:discussion}.

\section{Observations}
\label{sect:obs}

Observations were carried out with three long-baseline near-infrared
interferometers: IOTA, PTI, and VLTI.

IOTA, located on Mount Hopkins, was at the time of the observations a 
two-telescope stellar
interferometer  \citep{Tra98}, whose L-shape
multiple baseline configuration allows observations with baselines
ranging from 5 to 38 m, with limiting magnitudes of $K\sim 6, H\sim
6$.  PTI is a three-telescope interferometer located on Mount Palomar
\citep{Col99A} with limiting magnitudes of $K\sim 6.5, H\sim 5$ and
whose 3 baselines can be operated separately. The VLTI is an
interferometer that at the time of the observations could combine two
apertures among four $8$m unit telescopes (UTs) or two $0.4$m
siderostats \citep{Gli03}.  The VLTI was used with tip-tilt correction
on UT1 and UT3 and the VINCI focal instrument \citep{Ker00,Ker03}
which is a fiber-filtered instrument capable of combining two beams in
the $K$ band.

FU~Orionis was observed during five observing runs in 1998 with IOTA, in 1998,
1999, 2000, 2003 with PTI, and in 2002 with the VLTI.  We spent a total
of 42 nights on this project distributed into 29 nights on PTI,
11 on IOTA and 2 on VLTI. The distribution of these observations with time and
baselines is summarized in Table \ref{tab:obsdist}. A total of 287
observations have been performed leading to 287 fringe visibility points.

\begin{table}[t]
  \centering
  \caption{Distribution of observations over the campaign.}
  \label{tab:obsdist}
\def\multibrace#1{\multicolumn{#1}{c}{\upbracefill}}
\def\multientry#1#2{\multicolumn{#1}{c}{#2}}
\def\c#1{\multicolumn{1}{c}{#1}}
\begin{tabular}{lcrrrrrr}
   \hline \hline
               &     &\multientry{5}{\# nights}         &\c{\#}\\
                          \cline{3-7}
   Baseline    &Band &\c{98}&\c{99}&\c{00}&\c{02}&\c{03}&Data\\
   \hline
   IOTA/S15N15 &H    &   3  &      &      &      &      &    44 \\
   IOTA/S15N35 &H    &   5  &      &      &      &      &    39 \\
   PTI/NS      &H    &      &   2  &      &      &      &    13 \\
   PTI/NW      &H    &      &      &   2  &      &      &     4 \\[-1ex]
               &     &           \multibrace{5}        \\
   total       &     &         \multientry{5}{12}       &   100 \\
   \hline
   IOTA/S15N15 &K    &    1 &      &      &      &      &     4 \\
   IOTA/S15N35 &K    &    2 &      &      &      &      &    18 \\
   PTI/NS      &K    &   11 &   4  &    1 &      &    2 &   109 \\
   PTI/NW      &K    &      &      &    4 &      &      &    32 \\
      PTI/SW   &   K &      &      &      &      &    3 &    10 \\
   VLTI/U1-U3  &K    &      &      &      & 2    &      &    14 \\[-1ex]
               &     &           \multibrace{5}        \\
   total       &     &       \multientry{5}{   30}      &   187\\
   \hline
\end{tabular}

\end{table}

Table \ref{tab:obs} summarizes the main characteristics of our observations:
dates, baselines, filters and calibrators.

\begin{table*}[t]
   \centering 
   \caption{Logs of FU Ori observations.}
   \label{tab:obs}
\tabcolsep=.4em
\begin{tabular}{llllp{6cm}}
   \hline\hline
   \multicolumn{2}{l}{Observing period} &Baselines                        &Filters  & Calibrators \\ 
   \hline%
   1998-Nov-14     &  1998-Nov-27       & PTI 110m (NS)                   &K        &HD~42807, HD~37147, HD~38529, HD~32923\\
   1998-Dec-13     &  1998-Dec-26       & IOTA 21m (S15N15), 38m (S15N35) &K, H     &HD~42807, HD~37147, HD~38529, HD~31295\\
   1999-Nov-23     &  1999-Dec-01       & PTI 110m (NS)                   &K, H     &HD~42807, HD~37147, HD~38529, HD~32923\\
   2000-Nov-18     &  2000-Nov-27       & PTI 110m (NS), 85m (NW)         &K, H     &HD~42807, HD~37147, HD~38529, HD~32923\\
   2002-Oct-28     &  2000-Oct-29       & VLTI/VINCI 100m  (UT1-UT3)      &K        &HD~42807\\
      2003-Nov-19  &     2003-Nov-27    &    PTI 110m (NS), 87m (SW)      &   K     &HD~42807\\
   \hline
   \multicolumn{5}{p{0.9\hsize}}{%
      \small Angular diameters estimated from Hipparcos catalog in mas:\newline
      HD~42807: $0.5\pm0.1$, HD~37147: $0.3\pm0.2$, 
      HD~38529: $0.4\pm0.3$, HD~31295: $0.5\pm0.2$, 
      HD~32923: $0.75\pm0.30$.
   }\\
   \hline
\end{tabular}

\end{table*}

\section{Data processing}
\label{sect:dataproc}

The data processing involves two main steps: 
\begin{enumerate}
   \item The extraction of raw visibilities from recorded data, a procedure
      that is specific to each interferometer.
    \item The data calibration, i.e. dividing raw visibilities by an
      estimate of the instrumental transfer function, a procedure that
      is common to all data sets under the assumption that the raw
      visibilities have had all instrumental biases corrected by the
      previous step.
\end{enumerate}

\subsection{Extracting raw visibilities}

PTI data processing is described by \citet{Col99B}. The visibility is
based upon the ABCD algorithm which computes its estimate from
4 different phase quadratures of one fringe. Out of the four fringe
visibility estimation algorithms available in the standard PTI data
processing pipeline we chose the incoherent spectral estimator; the
spectrometer has a higher instrumental transfer function (due to the
spatial filtering effect of the single mode fiber), and the incoherent
estimator allows one to average visibilities over the entire H or K
bandpass without introducing biases due to atmospheric piston (cf.\
paper I).

The IOTA interferometer temporally encodes fringes.  We chose a
quadratic estimator similar to the one used for FLUOR \citep{Cou97}
with no photometric calibration signal. The two interferograms
simultaneously recorded at the output of the instrument are subtracted
to build a single interferogram with reduced photometric
contamination\footnote{Since the fringes are in phase
opposition, the energy in the spectral density distribution is
maximized at the fringe position and the photometric energy at lower
frequencies is reduced.}.  Visibilities are computed by estimating the
energy contained at the fringe position in the spectral power
density. We compute one average visibility from each batch of 500
interferograms and the standard deviation provides the error estimate
on this measurement.

The VINCI interferograms are, much like the IOTA case, also produced
by a temporal modulation of the optical path difference. The
processing applied to this signal is described in \citet{Ker04}, and
is based on a strategy comparable to that defined by \citet{Cou97},
with the optional use of wavelet transforms -- as opposed to Fourier
transforms -- for fringe detection and power spectrum computation. As
in the case of the IOTA data, we chose to use the visibilities
computed from the classical Fourier power spectrum method. Unlike the
IOTA case, we used the available photometric channels to compensate
for coupling efficiency variations. Each visibility measurement
results from the averaging of a 500-interferogram set of observations
and the bootstrapping method \citep{ET1993} is applied to derive its
associated internal error.

\subsection{Data external calibration}   

After the computation of raw visibilities for each interferometer, the
data reduction process enters the calibration path common to all the
interferometers. The key point is to estimate the system transfer
visibility for each observation of the source by observing calibration
sources for which one can estimate the fringe visibility that would be
measured with an unresolved source. These calibration sources
were chosen to minimize time- and sky-dependent variations, 
hence were close on the sky, and observed within 30 minutes of the science
target. The calibrators are corrected for their intrinsic visibility
loss due to their apparent diameter (inferred from Hipparcos distance
and size based on spectral type). We estimate the system transfer
visibility at the time of the target observations by interpolating the
calibrator diameter-corrected visibilities.  The weight of each
calibrator measurement is given by the inverse of the time delay
between the calibrator measurement and the target measurement.
Division of the observed target visibility with the interpolated
system visibility leads to the determination of the calibrated
visibility and its associated error.

\section{Results}
\label{sect:results}

\begin{figure*}[p]
   \centering 
   \includegraphics[width=0.65\hsize]{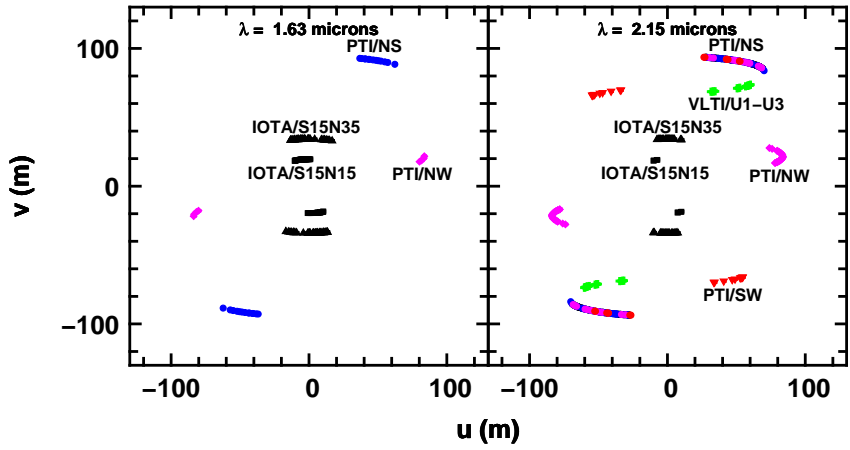}
   \caption{%
      (u,v) tracks corresponding to the observations of FU Orionis. 
      \strong{Left panel:} H data; \strong{right panel: K data.} The $(u,v)$
      points have been labeled with the interferometer and the baseline used.
   }
   \label{fig:uvcov}
\end{figure*}

\begin{figure*}[p]
   \centering
   \includegraphics[width=0.84\textwidth]{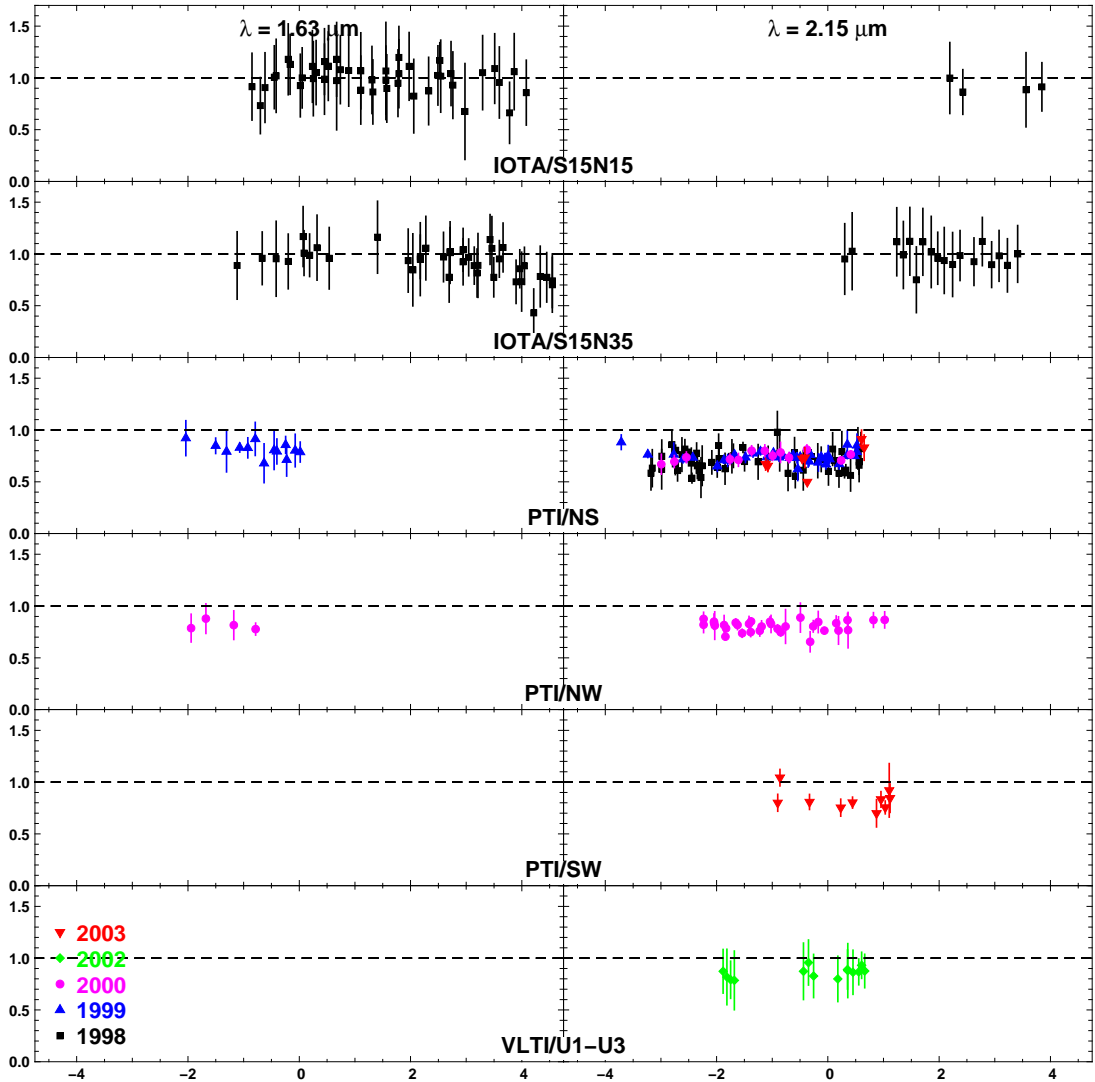}
   \caption{%
      Calibrated squared visibilities of FU
      Orionis as function of hour angle in H (left column) and K band
      (right column).  Rows from top to bottom display data obtained
      with the 6 interferometric baselines. 
   }
   \label{fig:ha} 
\end{figure*}

The observations with all interferometers covered a range of hour angles large
enough to allow the projected baselines on the sky to rotate over a significant
angle.  Figure \ref{fig:uvcov} displays the resulting $(u,v)$ plane coverage of
our observations with the six baselines: 21m (IOTA-S15N15), 38m
(IOTA-S15N35), 85m (PTI-NW), 87m (PTI-SW), 100m (VLTI-U1U3) and 110m
(PTI-NS).

Figure \ref{fig:ha} shows our results for the six baselines in the
H and K bands (left and right columns respectively). All panels show
calibrated squared visibilities as a function of hour angle. The main raw 
results are:
\begin{itemize}
   \item We confirm the 1997 result (paper~I): FU Orionis is resolved at 110m
      in the K band with the same visibility range (PTI-NS data) of
      $0.72\pm 0.07$ (cf. paper I). 
    \item FU Orionis is also resolved at about the same visibility
      level with the 85m PTI-NW, 87m PTI-SW, and 100m VLTI-U1-U3
      baselines.
   \item FU Orionis is unresolved at short baselines (IOTA data).
   \item We obtain color information between $1.6\microns$ (H-band) and
      $2.15\microns$ (K-band).
    \item We detect a peak-to-peak oscillation of 10\% at 110m in the
      K band and more marginally in the H band (this oscillation
      appears better in binned data of PTI/NS displayed in the right
      panel of Fig.~\ref{fig:binvis}).
\end{itemize}


\section{Interpretation of FU Ori observations}
\label{sect:interp}

Interpretation of the observations performed with infrared
long-baseline interferometers is not obvious since we have no direct
information about the brightness distribution of the object. However,
the measured visibilities allow us to set constraints on current
models. In paper~I, we had only one visibility point at 110m.
However, thanks to the spectral energy distribution of FU Ori, we
concluded that the data were compatible with a standard accretion disk
model. The binary scenario was also compatible with the data, but
still required an accretion disk to explain the SED.

We see here in the data the conjunction of 2 phenomena: a
  global decrease of the visibility with
  baseline and an oscillation as function of the spatial
  frequency. This can be explained by:
\begin{itemize}
\item the presence of a source which is marginally
  resolved and which dominates the overall flux,
\item and by the presence of a second, fainter source. The fainter
source can produce the oscillatory feature if it is located at a
distance larger than the typical size of the compact centered
structure.
\end{itemize}
The total complex visibility which results from these two objects is
composed of the addition of the two terms \citep[see][ for the
mathematical details]{Bod00}.

\subsection{Small-scale and bright structure}
\label{sec:small-scale}

In this section we focus our attention on the visibility behavior
without considering the oscillations.  Though we perform the model
fits with all individual measurements, for clarity we plot the results
with baseline-averaged visibilities, as displayed in
Table~\ref{tab:visavg}.

\begin{table}[t]
   \centering 
   \caption{Averaged square visibilities for the 6 baselines.}  
   \label{tab:visavg}
\tabcolsep=.4em
\begin{tabular}{lrrrr}
\hline\hline
       baseline &$\lambda$&    $B_\text{p}$ (m) &      $\theta$ ($\degree$) &         $|V^2|$\\
\hline
    IOTA/S15N15 &      H & $  20.2 \pm    0.5$ & $ 104.5 \pm    9.5$ & $1.00 \pm 0.13$\\
    IOTA/S15N35 &      H & $  35.0 \pm    0.7$ & $  95.7 \pm   13.7$ & $0.91 \pm 0.14$\\
         PTI/NS &      H & $ 103.7 \pm    2.2$ & $  61.8 \pm    3.5$ & $0.83 \pm 0.04$\\
         PTI/NW &      H & $  85.6 \pm    1.5$ & $  14.1 \pm    0.8$ & $0.79 \pm 0.05$\\
\hline
    IOTA/S15N15 &      K & $  20.8 \pm    0.3$ & $ 114.9 \pm    3.4$ & $0.90 \pm 0.13$\\
    IOTA/S15N35 &      K & $  34.4 \pm    0.3$ & $  91.4 \pm    7.8$ & $0.98 \pm 0.09$\\
         PTI/NS &      K & $ 102.7 \pm    3.8$ & $  63.6 \pm    6.5$ & $0.72 \pm 0.08$\\
         PTI/NW &      K & $  84.2 \pm    1.8$ & $  14.0 \pm    1.8$ & $0.79 \pm 0.05$\\
         PTI/SW &      K & $  82.5 \pm    2.9$ & $ -55.6 \pm    4.9$ & $0.82 \pm 0.08$\\
     VLTI/U1-U3 &      K & $  89.6 \pm    7.2$ & $  54.2 \pm    5.3$ & $0.87 \pm 0.05$\\
\hline
\end{tabular}

\end{table}

\subsubsection{Fitting strategy}

We use a disk model and a methodology similar to those of paper~I, but
the data and technique used have undergone several changes: the method
to compute visibilities is now based on Hankel transforms,
valid for axisymmetric distributions (See
Eq.~\ref{eq:V:face-on}); in the disk model the exponent of the
temperature profile power-law is a free parameter, allowing us to describe
flared irradiated disks as well as self-heated viscous ones;
visibility data taken with all baselines are included in the study;
the SED now comprises observations of the \citet[][ 5th
edition]{CIO99} catalog and from the 2MASS survey \citep{2mass}.
This model, although very simple, is able to reproduce a variety
  of situations without introducing new free parameters.

The model consists of a geometrically flat and optically
  thick disk with an inner radius {\rmin} and outer radius {\rmax}, and
features a radial temperature profile given by
\begin{equation}
   T(r) = \Tdisk \left( \frac r\rdisk \right) ^ {-\qdisk},
\end{equation}
where $\rdisk = 1$ AU is the reference radius, {\Tdisk} the effective
temperature at {\rdisk}, and {\qdisk} a parameter\footnote{See
  \citet{MB95,LMM03} for details on the morphology of disks and their
  relationship with the $q$ parameter.} usually ranging from $0.50$
(flared irradiated disks) to $0.75$ (standard viscous disks or
flat irradiated disks).  Each part of the surface of the disk located
at radius $r$ is assumed to emit as a blackbody at temperature $T(r)$,
so that the flux and visibility are deduced by radial integration for
a face-on disk with an inclination angle $\idisk=0$:
\begin{align}
  \flux(0)      &= \frac{2\pi}{d^2} \integ\rmin\rmax{r B_\lambda[T(r)]}{r}\\
  \vis(\Bp, 0)  &= \frac{1}{\flux(0)} \frac{2\pi}{d^2}
  \integ\rmin\rmax{ r B_\lambda[T(r)] J_0 \left[\frac{2\pi}{\lambda} \Bp \frac{r}{d} \right]}{r} \label{eq:V:face-on}
\end{align}
with $J_0$ the zeroth-order Bessel function of the first kind,
$B_\lambda$ the Planck function, $d$ the distance of the system and {\Bp} a  projected baseline.  
For a disk presenting an inclination $\idisk$ and a position angle 
$\theta$, these quantities become
\begin{align}
   \flux(\idisk)        &= \flux(0) \cos\idisk,\\
   \vis (\vecBp,\idisk) &= \vis \left( \Bp(\idisk,\theta), 0 \right),
\end{align}
where the corresponding baseline is 
\begin{align}
   \Bp(\idisk,\theta) &= \sqrt{B_{u,\theta}^2 + B_{v,\theta}^2 \cos^2 \idisk}
      \intertext{
and the coordinates are expressed in the reference frame
       rotated by the angle {\thetadisk},
      }
   B_{u,\theta}       &= B_u \, \cos\theta - B_v \, \sin\theta\\
   B_{v,\theta}       &= B_u \, \sin\theta + B_v \, \cos\theta.
\end{align}

The companion, FU~Ori~S, observed by \citet{WAHP04} and the
contribution of the FU~Ori central stellar source have
negligible and compensating effects on the measured visibilities.
FU~Ori~S is located $0.5\,\arcsec$ away from the bright
object, far outside the interferometric field of view (viz. the
``delay beam'', $\theta_{d}=\frac{\lambda^2}{\Delta\lambda B} \sim
0.1\,\arcsec$), and will therefore only combine incoherently with the flux
from FU~Ori~N.  The bias introduced by this incoherent flux
corresponds to the contribution of this companion flux to the total,
i.e. 1.6\% decrease in visibility. In addition, we neglect the
contribution of the central star, since it is unresolved and will only
slightly increase the visibilities. The contribution is of the order
of 2\%, almost cancelling the decrease of visibility due to FU~Ori~S.

\subsubsection{Best fit and parameter constraints}

\begin{table}[t]
   \centering
   \caption{%
      Model parameters for the small-scale structure.  This best fit
      has a reduced $\chi^2$ of 1.16 (with 306 individual measurements)
      and a ``goodness of fit'' of 2\%. The 3-$\sigma$ uncertainties on 
      each parameter have been determined from a $\chi^2$ map around 
      the minimum, as shown in left part of Fig.~\ref{fig:chi2}.
   }
   \label{tab:disk:param}
\begin{tabular}{llll}
\hline
\multicolumn{4}{c}{Accretion disc}\\
$\rmin$          & $    5.5_{-1.8}^{+2.9}$ $R_\odot$&
$\rmax$          & $    100$ AU (fixed)\\
$\Tdisk$         & $    745_{-24}^{+24}$ K&
$\qdisk$         & $   0.71_{-0.04}^{+0.05}$ \\
$\idisk$         & $     55_{-7}^{+5}\,\degree$&
$\thetadisk$     & $     47_{-11}^{\hphantom{0}+7}\,\degree$\\
\hline
\end{tabular}

\end{table}

\begin{figure*}[p]
   \centering
   \parindent=0pt
   \parbox{0.45\linewidth}{%
      \centering
      \includegraphics[width=0.95\hsize]{2556fig3a}\\
      \includegraphics[width=0.95\hsize]{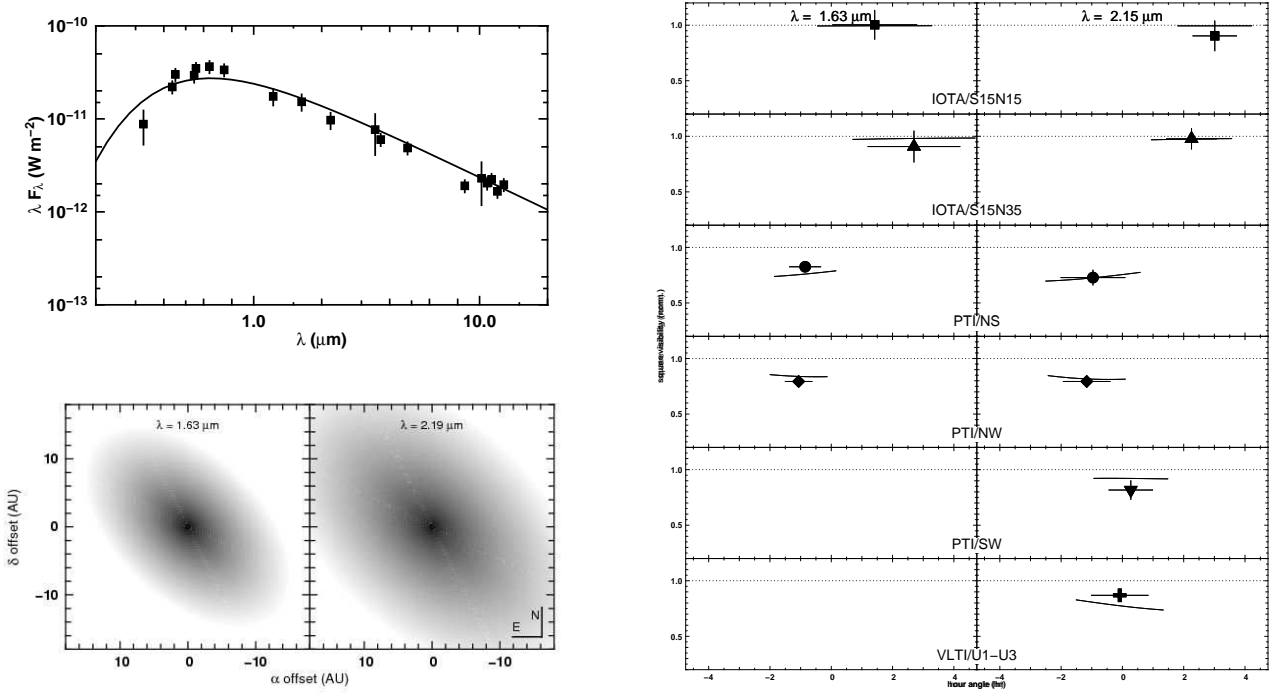}
   }%
   \parbox{0.55\linewidth}{%
      \centering
      \includegraphics[width=0.85\hsize]{2556fig3b}
   }
   \caption{%
     \emph{Accretion disk} model with baseline-averaged data corresponding to
     parameters listed in Table~\ref{tab:disk:param}. Lines correspond
     to the model and symbols to measurements.  The dotted line
     corresponds to an unresolved object. For the sake of clarity,
     data have been binned in the figure, but the model fit was
     performed using individual measurements. \strong{Top-left panel:}
     spectral energy distribution and best fit model.  \strong{Right
       panel:} visibility data versus hour angle in H and K.
     \strong{Bottom-left panel:} synthetic images in H and K in
      logarithmic scale. East is left and North is up.%
   }
   \label{fig:disk:fit}
\end{figure*}

\begin{figure*}[p]
   \centering
   \parindent=0pt
   \parbox{0.45\linewidth}{%
      \centering
      \includegraphics[width=0.95\hsize]{2556fig5a}\\
      \includegraphics[width=0.95\hsize]{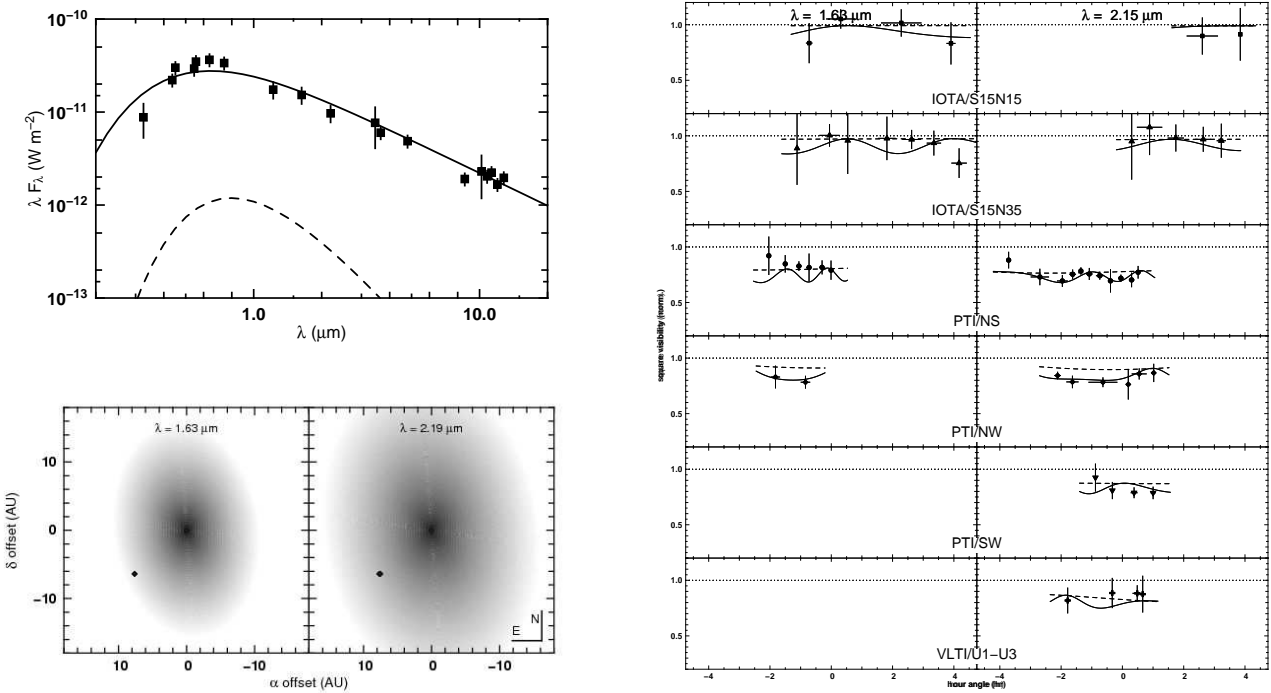}
   }%
   \parbox{0.55\linewidth}{%
      \centering
      \includegraphics[width=0.85\hsize]{2556fig5b}
   }
   \caption{
      \emph{Disk-spot} model corresponding to parameters of 
      Table~\ref{tab:binmod}. Lines corresponds to the best model fit 
      and markers to visibilities from Fig.~\ref{fig:ha} in
      binned form.
      \strong{Top-left panel:} spectral energy distribution and best
      fit model.  The solid line stands for the whole model and the
      dashed one for the spot.  
      \strong{Right panel:} visibility data vs. hour angle for each 
      baseline in H and K.  The contribution of the disk is displayed 
      in dashed lines.
      \strong{Bottom-left panel:} synthetic image in logarithmic scale.
      East is left and North is up.%
   }
   \label{fig:binvis}
\end{figure*}

\begin{figure*}[t]
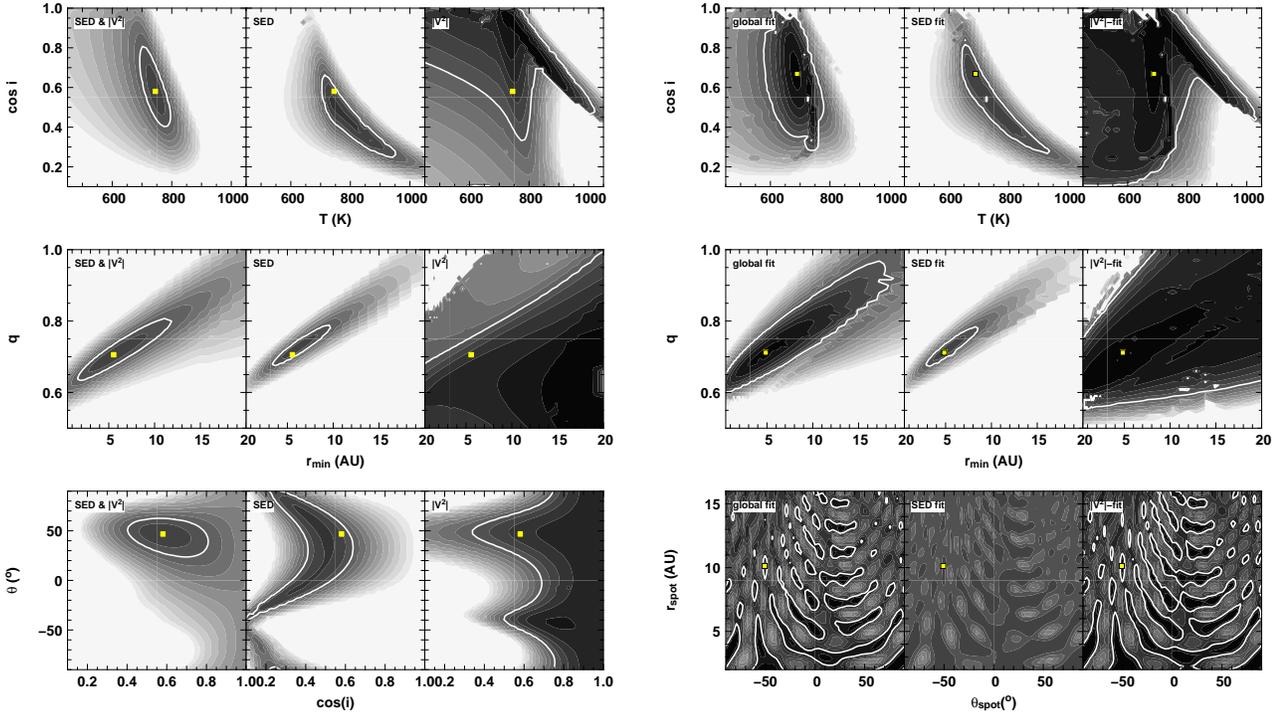

   \centering
   \begin{minipage}{0.48\hsize}
     \includegraphics[width=0.95\hsize]{2556fig4a}\\
     \includegraphics[width=0.95\hsize]{2556fig4b}\\
     \includegraphics[width=0.95\hsize]{2556fig4c}
   \end{minipage}
   \begin{minipage}{0.48\hsize}
     \includegraphics[width=0.95\hsize]{2556fig6a}\\
     \includegraphics[width=0.95\hsize]{2556fig6b}\\
     \includegraphics[width=0.95\hsize]{2556fig6c}
   \end{minipage}
 \caption{%
   $\chi^2$ distribution of the \emph{disk alone} (left column) and
   \emph{disk-spot} (right column) models around the best fit (marked
   with a square). The white solid line draws the 3-$\sigma$ limit
   (``goodness-of-fit'' $> 0.25$\%).
   \strong{Top:} inclination vs. temperature. \strong{Middle:}
   temperature law exponent vs. minimum radius. \strong{Bottom left:}
   inclination vs. position angle of the disk. \strong{Bottom right:}
   position angle of the spot vs. projected distance to the centre.
   \strong{Left subpanels:} $\chi^2$ determined with both visibility and
   SED data.  \strong{Middle subpanels:} $\chi^2$ determined with the SED
   only.  \strong{right subpanel:} $\chi^2$ determined with the
   visibilities only.}
   \label{fig:chi2}
\end{figure*}

When trying to interpret the measured visibilities and the SED
for the small-scale structure with the previous simple disk model
(neglecting the spot), the best fit has been unambiguously found
using a $\chi^2$-gradient search method.  The reduced $\chi^2$ is
1.16, and the probability that statistical deviation from the model
results in as large a deviation is 2\% (``goodness of fit'').  This
implies that the model is unlikely to explain the observations (since
we have been quite conservative in our determination of the data
uncertainty), but we cannot rule out this possibility. This low
\emph{goodness} gives credence to trying the more complicated model
with a spot (see Sect.~\ref{sect:disk+spot}).  The parameters
obtained are presented in Table~\ref{tab:disk:param}, with the
standard uncertainty on the parameters.  These errors have been
determined from the $\chi^2$ distribution displayed in left part of
Fig.~\ref{fig:chi2} --- assuming a Gaussian error distribution.  We
checked that the uncertainties derived have values close to those
predicted from the quadratic estimate of the $\chi^2$ near its
minimum.

Unlike interferometric observations of T~Tauri and Herbig Ae stars
\citep{Mil99,Ake00,Ake02,Eis03,Eis04}, the visibilities for FU~Ori are
compatible with the simple power-law radial temperature profile
predicted by the standard disk model. We derive a temperature
exponent $\qdisk \approx 0.71 \pm 0.05$, which is actually strongly
constrained by the slope of near-IR SED, and fairly consistent with
the visibilities.  The derived high temperature of about
$745\pm24\mbox{ K}$ at 1~AU hints to a viscous, self-heated disk
rather than an irradiated disk. 

The shape of the visible SED sets a constraint on the cut-off radius
$\rmin \approx 3$--$8\,R_\odot$.  We find that the intensity
of the near-IR emission strongly determines $\Tdisk \cos \idisk$,
while individual constraints on $\Tdisk$ and $\idisk$ are provided by
the visibilities.  The latter indeed determine {\Tdisk}, {\rmin},
{\idisk}  and {\thetadisk} (they have an influence resp. on
the spatial extent of the disk in H and K, on the amount of central
unresolved flux, and on the spatial extent in a particular direction),
but their impacts are coupled.  However, the visibilities also
constrain the radial brightness distribution, provided that
they be taken across a range of wavelengths, as explained by
\citet{Mal02,Lac03}\footnote{These authors find a value for the
temperature exponent $\qdisk \approx 0.6 \pm 0.1$ with a $1\sigma$
uncertainty, which is different from the value derived in these
papers.  This discrepancy seems to come from the reduced $(u,v)$
coverage used in that paper and from their ignoring the SED (which
pushes $\qdisk$ towards $0.75$).}.  The visibilities at different
baseline orientations constrain the inclination and the
position angle of the disk.  The visibilities at short baselines put
no constraints on the geometry of the disk because it is only
marginally resolved \citep{Lac03}, but we can rule out the
contribution of an extended (sub-arcsec) structure (e.g.  scattering
by an envelope) because IOTA visibilities at $\Bp \lesssim 30$\,m are
unresolved\footnote{An extended contribution yields a quick
decrease of the visibilities with baseline.}.

Figure \ref{fig:disk:fit} illustrates the comparison between the model
and the observations. The SED data and the result of the fit are
displayed in the upper left panel. The visibilities averaged by
baseline, listed in Table~\ref{tab:visavg}, are presented in the right
panel with the visibility curves corresponding to the $(u,v)$ track of
each baseline. Finally in the lower left panel, we show the
corresponding synthetic image of the disk.

\subsection{Large-scale structure}
\label{sect:disk+spot}

In this section we focus our attention on the high-frequency
oscillation in the visibility function, mostly visible in the 110m
$K$-band data of PTI.  We believe it to come from an
off-center unresolved light source that we call a ``spot''.
We fit to the data a compound source model consisting of an accretion
disk (as described above) and an unresolved spot.  The latter is
modeled as a circular black-body of uniform brightness with a
temperature $\Tspot$ and a radius $\rspot$.  The two components are
separated by the projected physical distance $\dsep$ with a position
angle $\thetasep$ (with a $180\,\degree$ ambiguity).  The fitting procedure
is the same as in Sect.~\ref{sec:small-scale}.

\subsubsection{Properties of the spot}
\label{sec:large-scale:prop}

\begin{table}[t]
   \centering
   \caption{%
      Model parameters for the large-scale structure without secular variation.  The reduced $\chi^2$ 
      of the best fit is 0.95 (with 306 individual measurements) and the 
      ``goodness of fit'' is 70\%.
   }
   \label{tab:binmod}
\begin{tabular}{llll}
\hline\hline
\multicolumn{4}{c}{Binary system}\\
      $\dsep$       \textsuperscript{a} & $    10.1_{-0.3}^{+0.4}$ AU 
     &$\thetasep$   \textsuperscript{a} & $    -50_{-1}^{+1}\,\degree$\\
\hline
\multicolumn{4}{c}{Accretion disc}\\
$\rmin$          & $    4.9_{-3.7}^{+2.8} $ $R_\odot$&
$\rmax$          & $    100$ AU (fixed)\\
$\Tdisk$         & $    692_{-41}^{+29}$ K&
$\qdisk$         & $   0.71_{-0.04}^{+0.08}$ \\
$\idisk$         & $     48_{-10}^{\hphantom{0}+9}\,\degree$&
$\thetadisk$\textsuperscript{b}     & $      8_{-21}^{+21}\,\degree$\\
\hline
\multicolumn{4}{c}{Uniform stellar disc}\\
$\Tspot$\textsuperscript{b}                & $   4600_{-800}^{+800}$ K&
$\rspot$                                   & $    5.0\ R_\odot$ (fixed)\\
\hline
\multicolumn{4}{l}{\textsuperscript{a} Other distant minima exist.}\\
\multicolumn{4}{l}{\textsuperscript{b} Quadratic error from the fitting routine}\\
\hline
\end{tabular}

\end{table}

The best model fit was obtained with the parameters listed in
Table~\ref{tab:binmod}. The reduced $\chi^2$ is 0.95 and the
probability that statistical deviation from the model accounts for as
large a deviation is 70\%, so the fit appears as a consistent
explanation for the data -- in comparison to the 2\% of the disk-alone
model -- but the detection of the structure is only at a 2-$\sigma$
level. The best-model fit is shown in Fig.~\ref{fig:binvis}, and the
probability distribution around the minimum is shown in the right part of
Fig.~\ref{fig:chi2}.

The disk parameters and their uncertainties are consistent with the
results of Sect.~\ref{sec:small-scale} except for the position angle:
its uncertainty is much larger in our last model fit ($8 \pm
20\,\degree$) than in the previous one ($47 \pm 10\,\degree$), and
both values differ by $2\sigma$.  This angle is constrained (together
with the inclination) by the visibility difference at three different
baseline angles, namely with the NW, SW, and NS PTI baselines (see
the right panel of Fig.~\ref{fig:binvis}).  When a
spot is added, this difference is changed, resulting in a
different position angle: in our best fit, the disk visibility for the
NS baseline is unchanged since it is the average of the oscillations,
while it is higher for the NW and SW baselines because the measured visibility
(disk plus spot) is in the low of the oscillations.

As Fig.~\ref{fig:chi2} shows (see bottom panel, $\chi^2$ as a
function of the spot location), there are multiple acceptable ``best'' model
fits featuring radically different spot locations.  The aforementioned
solution is actually the best in terms of modelling the PTI/NW and NS
oscillation in K, but was not found to be statistically better than others that
do not feature such oscillation.  We think, however, that the oscillation is
actual and that the model ambiguity results from the poor accuracy of other 
measurements (IOTA in H and K, and PTI in H).  Thus, results on the spot
location should be taken with caution.

We have chosen to parametrize the flux of this spot in terms of
stellar photosphere with the temperature and the radius of the
photosphere as parameters. However only the $K$ data can be used to
constrain them and we have fixed the value of the radius to $5\Rsun$,
which is typical for a young star. The derived temperature depends
strongly on the value taken for the radius and therefore should be
considered with caution. 

The location of the spot is quite well constrained by the shape
of the oscillation with a projected distance to the center of
$10\pm1$\,AU and a position angle $-50\pm1\,\degree$ (with 180 deg
ambiguity).  The amplitude of the oscillations in the K-band
constrains the flux ratio of the spot to the primary, $\Delta K
\approx3.9\pm0.2$ and $\Delta H \approx 3.6\pm0.2$\,mag.  However,
there are other acceptable $\chi^2$ minima (goodness-of-fit $\gtrsim
50$\%, see the bottom right $\chi^2$ maps in Fig.~\ref{fig:chi2})
which do not reproduce the oscillation, but they cannot be ruled out.
We should note that the oscillations are detected mostly in the $K$
band and therefore $\Delta H$ is not constrained by both the
oscillations and the model, since there is little chance that a spot
at $4000-5000\,\text{K}$ be a pure black-body.


\subsubsection{Secular evolution}

Assuming that the spot location of the disk-spot model is that derived
in Sect.~\ref{sec:large-scale:prop} (which allows us to explain the
PTI oscillation in K), we carried out a model fit of its location for
each year of observation. The results are displayed in
Table~\ref{tab:spot:move}.  The fit is successful in the years 1998,
1999, and 2000 and feature a 3-$\sigma$ detection of an almost radial
motion. The data obtained in 2003 are not sufficient to
  constrain the position of the spot for this year
  explaining the bad $\chi^2$.

\begin{table}[t]
   \centering
   \caption{%
      Parameters of the best model fit of the location of the spot in 1998,
      1999, 2000, and 2003, and its quality.%
   }   
   \label{tab:spot:move}
\begin{tabular}{lrrrr}
   \hline\hline
     year &           $\dsep$ (AU) &     $\thetasep$ ($\degree$) & $\chi^2$ (total) & $\chi^2$ ($|V|^2$)\\
   \hline
     1998 & $    8.0 \pm     1.9$ & $  -48.0 \pm     3.3$ &       0.63 &       0.54\\
     1999 & $    9.6 \pm     1.5$ & $  -49.4 \pm     2.2$ &       0.95 &       0.78\\
     2000 & $   10.1 \pm     0.6$ & $  -50.3 \pm     1.6$ &       1.04 &       0.91\\
     2003 & $   11.1 \pm     0.3$ & $  -49.9 \pm     1.7$ &       2.54 &       4.02\\
   \hline
\end{tabular}

\end{table}

\section{Discussion}
\label{sect:discussion}

\subsection{Accretion disk versus photosphere}

The current data set confirms the result from paper I and shows
consistent visibility data. We can definitively confirm that we are
observing an object whose size (Gaussian full width at half
  maximum) is 1.5mas in $K$ wide corresponding to 0.7AU or $150\Rsun$
at a distance of 450 pc (see paper I for details).  Therefore if
\citet{Her03} is correct in interpreting the high resolution
spectroscopic data as stellar chromospheric activity, then the
star must be accompanied by cooler circumstellar material that spreads
beyond $150\Rsun$ that is probably a circumstellar disk.  

The temperature $\Tdisk$ obtained in our fits yield an accretion
rate\footnote{Uncertainties include the fitting uncertainty and the
  uncertainty on the boundary condition in the standard model (given
  by the stellar radius, liberally assumed to be between 0.5 to 6
  solar radii).} of $\Mdot = (6.3 \pm 0.6) \times 10^{-5}
(\Mstar/\Msun)^{-1} \, \Msun/\yr$ for the disk alone and of $\Mdot =
(4.7 \pm 0.6) \times 10^{-5} (\Mstar/\Msun)^{-1} \, \Msun/\yr$ for the
disk accompanied by a spot.  The effective temperature-accretion rate
relation is the one of the standard model (see Eq. 2.7 in
\citealt{SS73}). It is interesting to note that the corresponding
temperature at the inner radius is 10050~K (resp.\ 10140~K). In our
present model, the disk is optically thick whatever the physical
process which dominates the opacity (gas or dust). A more detailed
study of this boundary region would be of great interest.

Compared to paper~I, we are able to begin constraining the
geometry of the accretion disk, with an inclination angle of the order
of 50 degrees and position angle of the order 10--40 degrees. More
accurate data and better $(u,v)$ coverage is still desirable,
but we already show the path toward such constraints.

Other observations of Herbig Ae stars \citep{Mil01,Ake02,Eis03,Eis04}
and of T Tauri stars \citep{Ake03,Col03} cannot be interpreted in the
framework of the standard disk model. Measured near-IR visibilities
are smaller than the theoretically expected values, suggesting
puffed-up inner disk walls, and flaring is often necessary to fit mid-
and far-IR photometry \citep{DDN01,Muz03}. In contrast, our work on FU
Ori, as well as some Herbig Be stars observed by \citet{Eis04}, shows
that the observed circumstellar disk structure is consistent with the
standard accretion disk model, i.e. geometrically flat and
optically thick with a temperature law index close to 0.75. We
suggest that accretion-dominated objects (such as FUors and Herbig Be
stars) fit the standard accretion disk model, while Herbig Ae stars
and T Tauris may have different disk structures due to the influence
of irradiation.  This is also consistent with the fact that in FUors
the disk completely dominates the luminosity and therefore the physics
of the disk would be expected to be governed by accretion and not by
external stellar heating.

\cite{HK85} have presented an interesting interpretation of double
lines observed in FU Ori by spectroscopy. The controversy with
\citet{Her03} comes from the fact that these authors are not able to
localize the origin of these lines. The explanation of \citet{HK85}
appears to be more consistent with our data.

The interferometric instrument AMBER on the VLTI \citep{Pet01} has the
potential to detect these double lines and identify the region where they are
emitted \citep{Mal03}.

\subsection{The nature of the bright spot}

\begin{figure}[tp]
   \centering
   \includegraphics[width=.95\hsize]{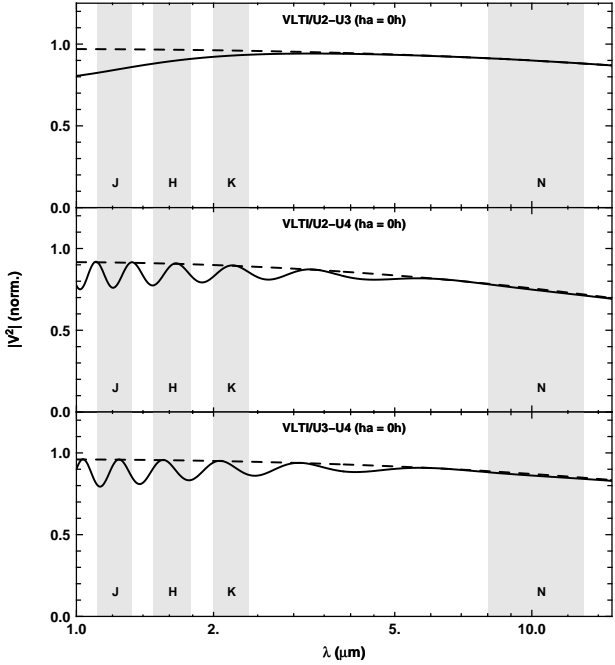}
   \caption{%
     Predicted squared visibility curve of the disk+spot model for
     FU~Ori through the infrared spectrum with the triplet of VLTI 8m
     telescopes U2-U3-U4.  }
   \label{fig:vlti}
\end{figure}

Our analysis of the data shows that the scenario which involves a
bright spot is more likely to explain our observations. The presence
of such a spot would modulate the visibility curves. Even if the
amplitude of the oscillations is not large and the number of cycles is
small, we believe that this spot is indeed present. As can be seen in
Fig.~\ref{fig:binvis}, the ripples of the visibility also explain some
low visibilities obtained at IOTA with short baselines.  This
detection was already pointed out by \cite{Ber00}, based on a smaller
data set, and the latest observations confirm this behavior. In
addition, our analysis by year, using independent sets of
observations, yields parameters which are close to each other. To
confirm the presence of this bright spot, we suggest observing FU
Orionis in several wavelengths and possibly with spectroscopy (see
Fig.~\ref{fig:vlti} for an example of the correlated spectrum that
would be obtained with a specific configuration of the VLTI with the
instruments AMBER and MIDI).

The nature of this bright spot is still unknown. This structure,
assuming it is unresolved, could be due to an increase of the
intensity in the disk due to a thermal outburst
\citep{CLP90,Bell94,Bell99}.  If the assumption of a photosphere
radius of $5\Rsun$ is correct (see remark in
Sect.~\ref{sec:large-scale:prop}), the derived color temperature of
$4600\,\text{K}$ of this spot is similar to temperatures observed in
stellar photospheres and therefore is consistent with the presence of
a young stellar companion.

The year-by-year analysis indicates that the spot may be moving with a
velocity of about $1.2\pm0.6 \,\AU\cdot\yr^{-1}$ on a rectilinear
trajectory. We have insufficient data to complete the analysis but
additional data will provide more time-sampling to confirm or exclude the
motion of the spot. The proper motion of FU Ori is $116\pm104 \times
10^{-5}\, {\rm s/yr}$ in right ascension and $86.4\pm15.5\, {\rm
mas/yr}$ in declination (Carlsberg Meridian Catalogue:
\cite{CMC}), i.e.  $39\pm7\,\AU/\yr$ at position angle of
$12\pm10\,\degree$. The spot motion, if actual, is therefore not due to a
foreground or background star. It could be explained by the motion of
a companion star orbiting on a very highly eccentric orbit around the
primary star or an orbit inclined with respect to the disk.

This explanation is consistent with the scenario introduced by
\citet{BB92} and recently revisited by \citet{BA04}
following the discovery of a binary companion to FU Ori
\citep{WAHP04}.  The companion observed by \citet{WAHP04} cannot be
responsible for, or be the consequence of, the 1936 outburst as noted
by \citet{BA04}.  However these same authors have also pointed out
that ``FU Orionis itself must be a close binary, with a semi-major
axis of $10\,\AU$ or less''. Confirmation and extension by
future measurements are needed to confirm that the spot we are
observing might indeed be the companion they were looking for located
on an inclined or highly eccentric orbit.

This spot could also represent the signature of the presence of a
young planet. Recently \citep{Lod04} proposed that FUor outbursts
could be caused by a planet at about 10$\Rsun$ which causes gap
instabilities. In the case of FU Ori, the mass of the planet would
have had to be $\sim 15\Mjup$ in order to explain the rapid rise of
the luminosity up to $500\Lsun$. We do not have enough color
measurements to derive the temperature and the luminosity of the spot
in order to check that it is compatible with a $15\Mjup$ planet. Nor
can we explain why the planet would be migrating away from the center.

\section{Conclusion}

We have presented the most complete set of interferometric
observations obtained so far in the domain of star formation with
several infrared interferometers: PTI, IOTA and VLTI. These
observations help us to better understand the exact
nature of FU Orionis. Our measurements confirm that FU Ori hosts an
active accretion disk with a $\approx 5\times10^{-5}\Msun/\yr$
accretion rate. We have also marginally detected an unresolved bright
structure, that we identify as a possible close companion on a highly
eccentric orbit. This putative companion may be responsible for the
outburst of the accretion rate that led to the rapid increase of the
luminosity.

These observations also show that the inner structure of YSOs is
rather complex and cannot be fully understood with only a small number
of interferometric observations. More wavelength and $(u,v)$ coverage
is required to fully constrain models. The next step will probably be
the reconstruction of an image of this interesting object based on extensive
observations with the VLTI.

\acknowledgements{This long term project would not have been possible
  without the support of many colleagues and funding agencies. We
  would like to thank the PTI collaboration for the opportunity to
  observe with PTI and especially C.~Koresko, C.~Beichman and
  S.~Kulkarni. We have welcome the help for observing at PTI of
  P.~M\`ege, P.~Kervella and K.~Rykoski. At IOTA, we have been helped
  by the IOTA consortium and also by P.~Haguenauer.  We wish to thank
  also the commissioning team of the VLTI who has participated in the
  collection of VLTI data. We are also grateful to the referee who
  helped us to clarify this paper for non interferometrists.  This
  work has made extensive use of the Internet services of the Centre
  de Donn\'ees de Strasbourg (SIMBAD, VizieR, Aladin) and of the
  Astronomical Database Services (ADS).  This project has benefited
  from funding from the French Centre National de la Recherche
  Scientifique (CNRS) through its INSU Programmes Nationaux (ASHRA,
  PNPS, ASGTE) and also from the partial financing support of the
  Laboratoire d'Astrophysique de Grenoble (LAOG). JAE acknowledges
  support from a Michelson Graduate Research Fellowship. BFL
  gratefully acknowledges the support of a Pappalardo Fellowhip in
  Physics. Finally we acknowledge numerous discussions with our
  colleagues in this field, namely B.~Reipurth and the members of the
  star and planet formation group at LAOG.}
  


\begin{thebibliography}{52}
\expandafter\ifx\csname natexlab\endcsname\relax\def\natexlab#1{#1}\fi

\bibitem[{{Akeson} {et~al.}(2003){Akeson}, {Ciardi}, \& {van Belle}}]{Ake03}
{Akeson}, R.~L., {Ciardi}, D., \& {van Belle}, G.~T. 2003, in Interferometry
  for Optical Astronomy II. Edited by Wesley A. Traub . Proceedings of the
  SPIE, Volume 4838, pp. 1037-1042 (2003)., 1037--1042

\bibitem[{{Akeson} {et~al.}(2002){Akeson}, {Ciardi}, {van Belle}, \&
  {Creech-Eakman}}]{Ake02}
{Akeson}, R.~L., {Ciardi}, D.~R., {van Belle}, G.~T., \& {Creech-Eakman}, M.~J.
  2002, \apj, 566, 1124

\bibitem[{{Akeson} {et~al.}(2000){Akeson}, {Ciardi}, {van Belle},
  {Creech-Eakman}, \& {Lada}}]{Ake00}
{Akeson}, R.~L., {Ciardi}, D.~R., {van Belle}, G.~T., {Creech-Eakman}, M.~J.,
  \& {Lada}, E.~A. 2000, \apj, 543, 313

\bibitem[{{Ambartsumian}(1971)}]{Amb71}
{Ambartsumian}, B.~A. 1971, Astrofizika, 7, 557

\bibitem[{{Ambartsumian} \& {Mirzoian}(1982)}]{Amb82}
{Ambartsumian}, V.~A. \& {Mirzoian}, L.~V. 1982, \apss, 84, 317

\bibitem[{{Bell}(1999)}]{Bell99}
{Bell}, K.~R. 1999, \apj, 526, 411

\bibitem[{{Bell} \& {Lin}(1994)}]{Bell94}
{Bell}, K.~R. \& {Lin}, D.~N.~C. 1994, \apj, 427, 987

\bibitem[{{Berger} {et~al.}(2000){Berger}, {Malbet}, {Colavita}, {Segransan},
  {Millan-Gabet}, \& {Traub}}]{Ber00}
{Berger}, J., {Malbet}, F., {Colavita}, M.~M., {et~al.} 2000, in Proc. SPIE
  Vol. 4006, p. 597-604, Interferometry in Optical Astronomy, Pierre J. L\'ena;
  Andreas Quirrenbach; Eds., 597--604

\bibitem[{{Boden}(2000)}]{Bod00}
{Boden}, A.~F. 2000, in Principles of Long Baseline Stellar Interferometry, ed.
  {Peter R. Lawson} ({JPL publication 00-009 07/00}), 25

\bibitem[{{Bonnell} \& {Bastien}(1992)}]{BB92}
{Bonnell}, I. \& {Bastien}, P. 1992, \apjl, 401, L31

\bibitem[{{Clarke} {et~al.}(1990){Clarke}, {Lin}, \& {Pringle}}]{CLP90}
{Clarke}, C.~J., {Lin}, D.~N.~C., \& {Pringle}, J.~E. 1990, \mnras, 242, 439

\bibitem[{{Clarke} \& {Syer}(1996)}]{CS96}
{Clarke}, C.~J. \& {Syer}, D. 1996, \mnras, 278, L23

\bibitem[{{Colavita} {et~al.}(2003){Colavita}, {Akeson}, {Wizinowich}, {Shao},
  {Acton}, {Beletic}, {Bell}, {Berlin}, {Boden}, {Booth}, {Boutell}, {Chaffee},
  {Chan}, {Chock}, {Cohen}, {Crawford}, {Creech-Eakman}, {Eychaner},
  {Felizardo}, {Gathright}, {Hardy}, {Henderson}, {Herstein}, {Hess},
  {Hovland}, {Hrynevych}, {Johnson}, {Kelley}, {Kendrick}, {Koresko}, {Kurpis},
  {Le Mignant}, {Lewis}, {Ligon}, {Lupton}, {McBride}, {Mennesson},
  {Millan-Gabet}, {Monnier}, {Moore}, {Nance}, {Neyman}, {Niessner}, {Palmer},
  {Reder}, {Rudeen}, {Saloga}, {Sargent}, {Serabyn}, {Smythe}, {Stomski},
  {Summers}, {Swain}, {Swanson}, {Thompson}, {Tsubota}, {Tumminello}, {van
  Belle}, {Vasisht}, {Vause}, {Walker}, {Wallace}, \& {Wehmeier}}]{Col03}
{Colavita}, M., {Akeson}, R., {Wizinowich}, P., {et~al.} 2003, \apjl, 592, L83

\bibitem[{{Colavita}(1999)}]{Col99B}
{Colavita}, M.~M. 1999, \pasp, 111, 111

\bibitem[{{Colavita} {et~al.}(1999){Colavita}, {Wallace}, {Hines}, {Gursel},
  {Malbet}, {Palmer}, {Pan}, {Shao}, {Yu}, {Boden}, {Dumont}, {Gubler},
  {Koresko}, {Kulkarni}, {Lane}, {Mobley}, \& {van Belle}}]{Col99A}
{Colavita}, M.~M., {Wallace}, J.~K., {Hines}, B.~E., {et~al.} 1999, \apj, 510,
  505

\bibitem[{{Copenhagen University Observatory} {et~al.}(1999){Copenhagen
  University Observatory}, {Royal Greenwich Observatory}, \& {Instituto Y
  Observatory de Marina}}]{CMC}
{Copenhagen University Observatory}, {Royal Greenwich Observatory}, \&
  {Instituto Y Observatory de Marina}. 1999, {Carlsberg meridian catalogue La
  Palma} (Carlsberg meridian catalogue La Palma / Copenhagen University
  Observaory, Royal Greenwich Observatory, and Instituto y Observatory de
  Marina.~San Fernando, Spain: Instituto y Observatorio de Marina, 1999.)

\bibitem[{{Coud\'e Du Foresto} {et~al.}(1997){Coud\'e Du Foresto}, {Ridgway},
  \& {Mariotti}}]{Cou97}
{Coud\'e Du Foresto}, V., {Ridgway}, S., \& {Mariotti}, J.-M. 1997, \aaps, 121,
  379

\bibitem[{{Cutri} {et~al.}(2003){Cutri}, {Skrutskie}, {van Dyk}, {Beichman},
  {Carpenter}, {Chester}, {Cambresy}, {Evans}, {Fowler}, {Gizis}, {Howard},
  {Huchra}, {Jarrett}, {Kopan}, {Kirkpatrick}, {Light}, {Marsh}, {McCallon},
  {Schneider}, {Stiening}, {Sykes}, {Weinberg}, {Wheaton}, {Wheelock}, \&
  {Zacarias}}]{2mass}
{Cutri}, R.~M., {Skrutskie}, M.~F., {van Dyk}, S., {et~al.} 2003, VizieR Online
  Data Catalog, 2246

\bibitem[{{Dullemond} {et~al.}(2001){Dullemond}, {Dominik}, \& {Natta}}]{DDN01}
{Dullemond}, C.~P., {Dominik}, C., \& {Natta}, A. 2001, \apj, 560, 957

\bibitem[{Efron \& Tibshirani(1993)}]{ET1993}
Efron, B. \& Tibshirani, R.~J. 1993, An Introduction to the Bootstrap,
  Monographs on Statistics and Applied Probability (Chapman \& Hall/CRC)

\bibitem[{{Eisner} {et~al.}(2003){Eisner}, {Lane}, {Akeson}, {Hillenbrand}, \&
  {Sargent}}]{Eis03}
{Eisner}, J.~A., {Lane}, B.~F., {Akeson}, R.~L., {Hillenbrand}, L.~A., \&
  {Sargent}, A.~I. 2003, \apj, 588, 360

\bibitem[{{Eisner} {et~al.}(2004){Eisner}, {Lane}, {Hillenbrand}, {Akeson}, \&
  {Sargent}}]{Eis04}
{Eisner}, J.~A., {Lane}, B.~F., {Hillenbrand}, L.~A., {Akeson}, R.~L., \&
  {Sargent}, A.~I. 2004, \apj, 613, 1049

\bibitem[{{Gezari} {et~al.}(1999){Gezari}, {Schmitz}, {Pitts}, \&
  {Mead}}]{CIO99}
{Gezari}, D.~Y., {Schmitz}, M., {Pitts}, P.~S., \& {Mead}, J.~M. 1999, {Catalog
  of infrared observations, fifth edition} (CDS-ADS)

\bibitem[{{Glindemann} {et~al.}(2003){Glindemann}, {Algomedo}, {Amestica},
  {Ballester}, {Bauvir}, {Bugue{\~ n}o}, {Correia}, {Delgado}, {Delplancke},
  {Derie}, {Duhoux}, {di Folco}, {Gennai}, {Gilli}, {Giordano}, {Gitton},
  {Guisard}, {Housen}, {Huxley}, {Kervella}, {Kiekebusch}, {Koehler}, {L{\'
  e}v{\^ e}que}, {Longinotti}, {M{\' e}nardi}, {Morel}, {Paresce}, {Phan Duc},
  {Richichi}, {Sch{\" o}ller}, {Tarenghi}, {Wallander}, {Wittkowski}, \&
  {Wilhelm}}]{Gli03}
{Glindemann}, A., {Algomedo}, J., {Amestica}, R., {et~al.} 2003, \apss, 286, 35

\bibitem[{{Hartmann} {et~al.}(2004){Hartmann}, {Hinkle}, \& {Calvet}}]{HHC04}
{Hartmann}, L., {Hinkle}, K., \& {Calvet}, N. 2004, \apj, 609, 906

\bibitem[{{Hartmann} \& {Kenyon}(1985)}]{HK85}
{Hartmann}, L. \& {Kenyon}, S.~J. 1985, \apj, 299, 462

\bibitem[{{Hartmann} \& {Kenyon}(1996)}]{HK96}
{Hartmann}, L. \& {Kenyon}, S.~J. 1996, Annual Review of Astronomy and
  Astrophysics, 34, 207

\bibitem[{{Herbig}(1977)}]{Her1977}
{Herbig}, G.~H. 1977, \apj, 217, 693

\bibitem[{{Herbig} {et~al.}(2003){Herbig}, {Petrov}, \& {Duemmler}}]{Her03}
{Herbig}, G.~H., {Petrov}, P.~P., \& {Duemmler}, R. 2003, \apj, 595, 384

\bibitem[{{Kenyon} {et~al.}(2000){Kenyon}, {Kolotilov}, {Ibragimov}, \&
  {Mattei}}]{Ken2000}
{Kenyon}, S.~J., {Kolotilov}, E.~A., {Ibragimov}, M.~A., \& {Mattei}, J.~A.
  2000, \apj, 531, 1028

\bibitem[{{Kervella} {et~al.}(2000){Kervella}, {Coud\'e du Foresto},
  {Glindemann}, \& {Hof\-mann}}]{Ker00}
{Kervella}, P., {Coud\'e du Foresto}, V., {Glindemann}, A., \& {Hof\-mann}, R.
  2000, in Proc. SPIE Vol. 4006, p. 31-42, Interferometry in Optical Astronomy,
  Pierre J. L\'ena; Andreas Quirrenbach; Eds., 31--42

\bibitem[{{Kervella} {et~al.}(2003){Kervella}, {Gitton}, {Segransan}, {di
  Folco}, {Kern}, {Kiekebusch}, {Duc}, {Longinotti}, {Coud\'e du Foresto},
  {Ballester}, {Sabet}, {Cotton}, {Schoeller}, \& {Wilhelm}}]{Ker03}
{Kervella}, P., {Gitton}, P.~B., {Segransan}, D., {et~al.} 2003, in
  Interferometry for Optical Astronomy II. Edited by Wesley A. Traub .
  Proceedings of the SPIE, Volume 4838, pp. 858-869 (2003)., 858--869

\bibitem[{{Kervella} {et~al.}(2004){Kervella}, {S\'egransan}, \& {Coud\'e du
  Foresto}}]{Ker04}
{Kervella}, P., {S\'egransan}, D., \& {Coud\'e du Foresto}, V. 2004, \aap, 425,
  1161

\bibitem[{{Kley} \& {Lin}(1999)}]{KL99}
{Kley}, W. \& {Lin}, D.~N.~C. 1999, \apj, 518, 833

\bibitem[{{Lachaume}(2003)}]{Lac03}
{Lachaume}, R. 2003, \aap, 400, 795

\bibitem[{{Lachaume} {et~al.}(2003){Lachaume}, {Malbet}, \& {Monin}}]{LMM03}
{Lachaume}, R., {Malbet}, F., \& {Monin}, J.-L. 2003, \aap, 400, 185

\bibitem[{{Larson}(1980)}]{Lar80}
{Larson}, R.~B. 1980, \mnras, 190, 321

\bibitem[{{Lodato} \& {Clarke}(2004)}]{Lod04}
{Lodato}, G. \& {Clarke}, C.~J. 2004, \mnras, 353, 841

\bibitem[{{Malbet} \& {Berger}(2002)}]{Mal02}
{Malbet}, F. \& {Berger}, J.-P. 2002, in {SF2A - Scientific Highlights 2001},
  ed. {F.~Combes, D.~Barret and F.~Th\'evenin} (EDP Sciences), 457--460

\bibitem[{{Malbet} {et~al.}(1998){Malbet}, {Berger}, {Colavita}, {Koresko},
  {Beichman}, {Boden}, {Kulkarni}, {Lane}, {Mobley}, {Pan}, {Shao}, {van
  Belle}, \& {Wallace}}]{Mal98}
{Malbet}, F., {Berger}, J.-P., {Colavita}, M.~M., {et~al.} 1998, \apj, 507,
  149, (paper I)

\bibitem[{{Malbet} \& {Bertout}(1995)}]{MB95}
{Malbet}, F. \& {Bertout}, C. 1995, \aaps, 113, 369

\bibitem[{{Malbet} {et~al.}(2003){Malbet}, {Bloecker}, {Foy}, {Fraix-Burnet},
  {Mathias}, {Marconi}, {Monin}, {Petrov}, {Stee}, {Testi}, \&
  {Weigelt}}]{Mal03}
{Malbet}, F., {Bloecker}, T., {Foy}, R., {et~al.} 2003, in Interferometry for
  Optical Astronomy II. Edited by Wesley A. Traub . Proceedings of the SPIE,
  Volume 4838, pp. 917-923 (2003)., 917--923

\bibitem[{{Millan-Gabet} {et~al.}(2001){Millan-Gabet}, {Schloerb}, \&
  {Traub}}]{Mil01}
{Millan-Gabet}, R., {Schloerb}, F.~P., \& {Traub}, W.~A. 2001, \apj, 546, 358

\bibitem[{{Millan-Gabet} {et~al.}(1999){Millan-Gabet}, {Schloerb}, {Traub},
  {Malbet}, {Berger}, \& {Bregman}}]{Mil99}
{Millan-Gabet}, R., {Schloerb}, F.~P., {Traub}, W.~A., {et~al.} 1999, \apjl,
  513, 131

\bibitem[{{Muzerolle} {et~al.}(2003){Muzerolle}, {Calvet}, {Hartmann}, \&
  {D'Alessio}}]{Muz03}
{Muzerolle}, J., {Calvet}, N., {Hartmann}, L., \& {D'Alessio}, P. 2003, \apjl,
  597, L149

\bibitem[{{Petrov} \& {Herbig}(1992)}]{PH92}
{Petrov}, P.~P. \& {Herbig}, G.~H. 1992, \apj, 392, 209

\bibitem[{{Petrov} {et~al.}(2001){Petrov}, {Malbet}, {Richichi}, {Hofmann},
  {Mourard}, {et~al.}}]{Pet01}
{Petrov}, R., {Malbet}, F., {Richichi}, A., {et~al.} 2001, C. R. Acad. Sci.
  Paris, t, 67

\bibitem[{{Reipurth} \& {Aspin}(2004{\natexlab{a}})}]{RA2004}
{Reipurth}, B. \& {Aspin}, C. 2004{\natexlab{a}}, \apjl, 608, L65

\bibitem[{{Reipurth} \& {Aspin}(2004{\natexlab{b}})}]{BA04}
{Reipurth}, B. \& {Aspin}, C. 2004{\natexlab{b}}, \apjl, 608, L65

\bibitem[{{Shakura} \& {Sunyaev}(1973)}]{SS73}
{Shakura}, N.~I. \& {Sunyaev}, R.~A. 1973, \aap, 24, 337

\bibitem[{{Traub}(1998)}]{Tra98}
{Traub}, W.~A. 1998, in Proc. SPIE Vol. 3350, p. 848-855, Astronomical
  Interferometry, Robert D. Reasenberg; Ed., 848--855

\bibitem[{{Wang} {et~al.}(2004){Wang}, {Apai}, {Henning}, \&
  {Pascucci}}]{WAHP04}
{Wang}, H., {Apai}, D., {Henning}, T., \& {Pascucci}, I. 2004, \aap, 601, L83

\end{thebibliography}

\end{document}